\documentclass[12pt]{iopart}

\usepackage{dcolumn}
\usepackage{amssymb}
\pdfoutput=1
\expandafter\let\csname equation*\endcsname\relax
\expandafter\let\csname endequation*\endcsname\relax
\usepackage{amsmath}
\usepackage{graphicx}
\usepackage{color}

\begin{document}

\title{Financial interaction networks inferred from traded volumes}

\author{Hongli Zeng$^{1}$, R\'emi Lemoy$^{1,2}$ and Mikko Alava$^{1}$}


\address{Aalto University,\\ 
$^{1}$Department of Applied Physics,\\
$^{2}$Department of Information and Computer Science,\\
 Espoo, Finland}

\eads{\mailto{hong.zeng@aalto.fi}, \mailto{remilemoy@gmail.com}, \mailto{mikko.alava@aalto.fi}}

\date{\today}


\begin{abstract}
In order to use the advanced inference techniques available for Ising models, we transform complex data (real vectors) into binary strings, by local averaging and thresholding. This transformation introduces parameters, which must be varied to characterize the behaviour of the system. 
The approach is illustrated on financial data, using three inference methods -- equilibrium, synchronous and asynchronous inference -- to construct functional connections between stocks. We show that the traded volume information is enough to obtain well known results about financial markets, which use however the presumably richer price information: collective behaviour ("market mode") and strong interactions within industry sectors. Synchronous and asynchronous Ising inference methods give results which are coherent with equilibrium ones, and more detailed since the obtained interaction networks are directed.
\end{abstract}

\section*{Introduction}
Inferring Ising models is an interesting and important field in the study of complex systems. Growing amounts of data are generated in different domains studying complex interacting systems: biology, economics, finance, social sciences and others. Inverse Ising problems provide very flexible tools which guess the interaction patterns from the observed data, even for non-equilibrium systems \cite{roudi2011,mezard2011,mastromatteo_marsiliJSTAT2011,tyrcha2013}, by linking pairwise correlations in the data to Ising couplings of the inferred model. An important question, which is studied in the present work, concerns the mapping of this complex data into binary variables. The data studied in inverse Ising problems consists indeed in binary strings, which explains why these models have been preferably applied to systems presenting activity/inactivity patterns, like neural networks \cite{schneidman2006Nautre, shlens2006Neuroscience}. 

But we show here how these tools can be applied to systems which provide more detailed data than activity/inactivity. This study deals with financial data generated by transactions on the New York Stock Exchange (NYSE). The activity of 100 highly traded stocks is recorded over a few years, and each trade is characterized by a time, a volume traded, and a price. Here we do not consider the price information and focus on time and volume traded. The simplest way to map such information to binary variables consists in forgetting about the traded volumes, and considering a trade as an activity, in the spirit of works on neural networks. This is done for instance by \cite{mastromatteo_marsiliJSTAT2011}, a work on which we build. However, in such work, the mapping of the data to binary strings has only one parameter, the length of the window. For large window sizes, one usually has a magnetization going to 1 -- as at least one trade is quite sure to happen if the window is large enough -- and low connected correlations. Then, in order to access phenomena taking place on bigger time scales, we use here a different mapping, which also allows us to consider volume information instead of just activity.

The paper presents the data, mapping and inference methods in section \ref{data}, the obtained couplings in section \ref{results}, and describes the inferred financial networks in section \ref{networks} before concluding.

\section{Data and inference}
\label{data}
We focus on the trades between 100 highly traded stocks of the NYSE, for 100 trading days between the 01.02.2003 and 30.05.2003. As \cite{mastromatteo_marsiliJSTAT2011}, we only study the $10^4$ central seconds of each day so that the system is not perturbed by the opening and closing periods of the stock exchange. Nonetheless, a Fourier transform of the data still shows a sharp peak at a frequency corresponding to this $10^4$ seconds long day -- a result also observed in \cite{mathiesen2013}.

This work has two particular details in the way it maps multi-dimensional financial trade data to binary strings of activity and inactivity. The first one is the use of a sliding time window, of length $\Delta t$, and shifted by a constant $\Delta s =1$ second, which is the time resolution of the data. This means that the information contained in two mapped datapoints separated by a time less than $\Delta t$ is partly redundant. In the case of a simple activity / inactivity mapping (neuronal data for instance), it also means that no information from the original data is lost. 

\begin{figure}[h!]
      \centering
        \includegraphics[width=10cm]{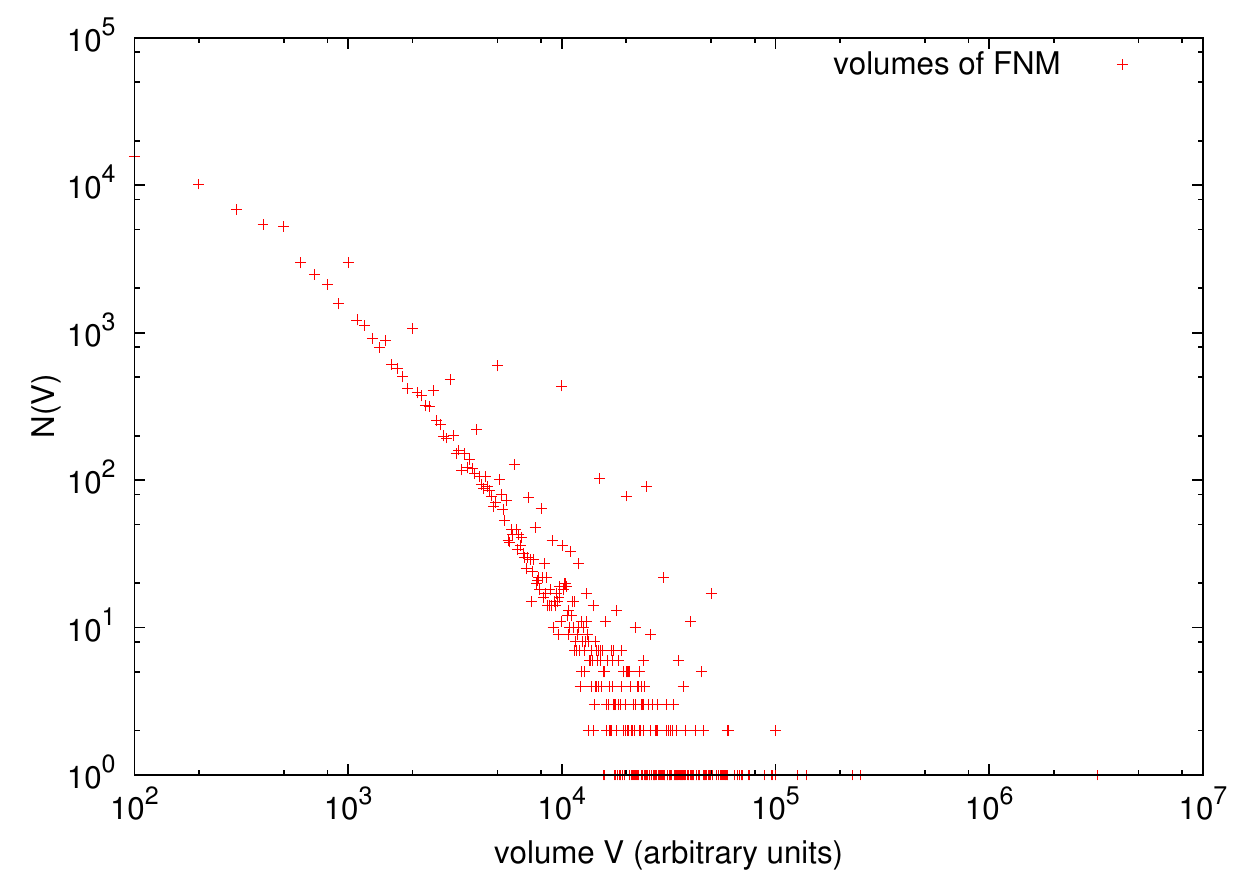}
	\includegraphics[width=10cm]{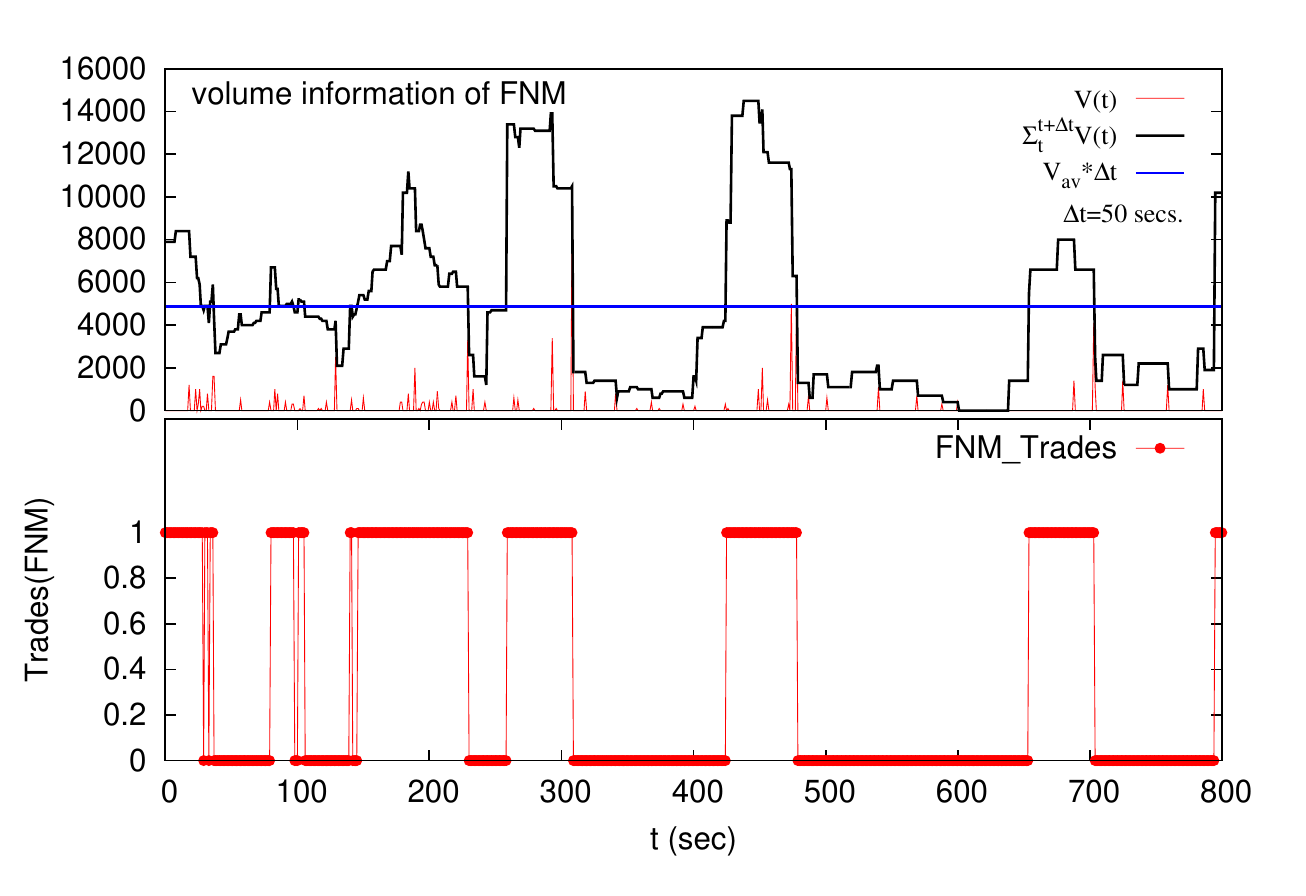}
        \caption{Upper panel: distribution of the volumes traded for the stock of Fannie Mae (ticker FNM), a mortgage investment company; Bottom panel: snapshot of the time series of traded volumes $V_i(t)$, summed volumes $V_i(t,\Delta t)$ and mapped trades of FNM, with $\Delta t =50$ seconds and $\chi=1$, which means that $V_{th}^i=V_{av}^i\Delta t$.}
	\label{fig1}
      \end{figure}

The second characteristic of our mapping is more important, and allows us to consider also the volume information related to a trade. This is motivated by the fact that the distribution of traded volumes in the data is very broad, as illustrated by the top panel of figure \ref{fig1}, where the distribution looks roughly like a power-law, and spreads over more than 3 decades (see also \cite{mathiesen2013} for distributions of trading volume rates). Our goal is to capture this important volume information through our mapping.
To this end, we consider for each stock $i$ the sum $V_i(t,\Delta t)$ of the volumes traded in the time window of length $\Delta t$ beginning at time $t$, and compare it to a given volume threshold $V_{th}^i=\chi V_{av}^i\Delta t$, where $V_{av}^i$ is the average volume of the considered stock traded per second\footnote{Note that $V_{av}^i$ is related to the average of the traded volumes histogram presented on the top panel of figure \ref{fig1} only through the average frequency of trades, which we do not study here.}, and $\chi$ a parameter governing our volume threshold:
\begin{equation}
   \rm s_i(t) = \left\{ \begin{array}{cl}
   1 & \mathrm{if}~~V_i(t,\Delta t) \geq V^i_{th}  \\
   -1 & \mathrm{if}~~V_i(t,\Delta t) < V^i_{th} \end{array} \right.
\end{equation}
This mapping, illustrated on the bottom panel of figure \ref{fig1}, ensures that volume information (and not only trade / no trade information) is taken into account in the obtained binary strings.

For instance with $\chi=1$, there will be a signal if the total volume traded in the considered time window exceeds the average. In addition, it preserves the possibility to use only trade / no trade information: if $\chi\rightarrow0$ (concretely, if $\chi$ is small enough that $V_{th}^i$ is for every stock $i$ smaller than the minimal traded volume, which is $100$ in the data), we obtain the usual trade / no trade pattern used for instance by \cite{mastromatteo_marsiliJSTAT2011}. 

Using this mapping, parameters $\Delta t$ and $\chi$ control the time and volume scales which will be explored by the inference. However, in order to completely characterize the behaviour of the system, the inference should be done for all possible values of these parameters $\Delta t$ and $\chi$, giving different results each time. The methods used (equilibrium, synchronous and asynchronous Ising inference) are described in \ref{inference}. In order to infer Ising couplings, equilibrium inference \cite{Kappen1998} uses equal-time pairwise correlations of the data. Synchronous \cite{roudi2011} and asynchronous inference \cite{zeng2013}, which work both on non-equilibrium systems, consider also respectively time-lagged pairwise correlations and their derivative with respect to the time lag.

\section{Results}
\label{results}
\begin{figure}[h!]
      \centering
        \includegraphics[width=10cm]{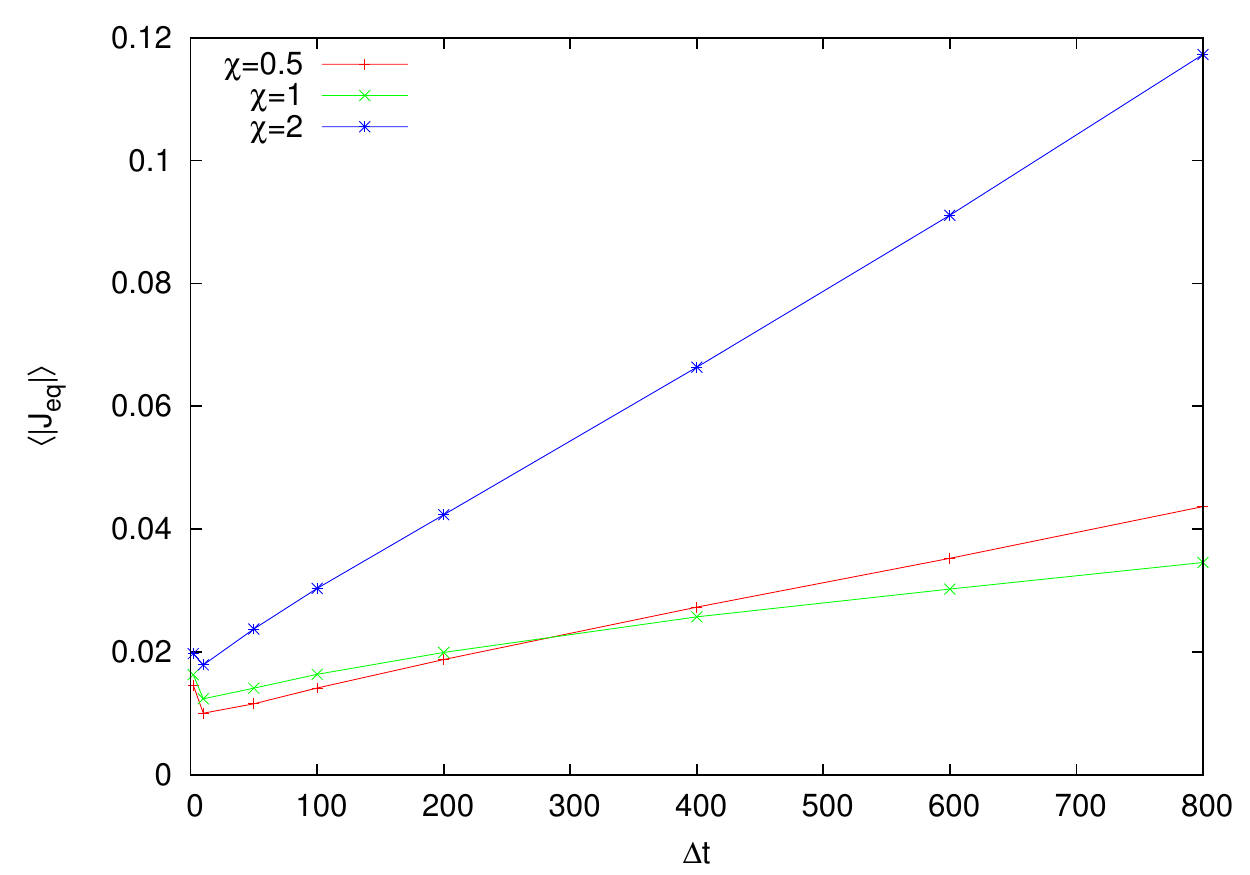}
	\includegraphics[width=10cm]{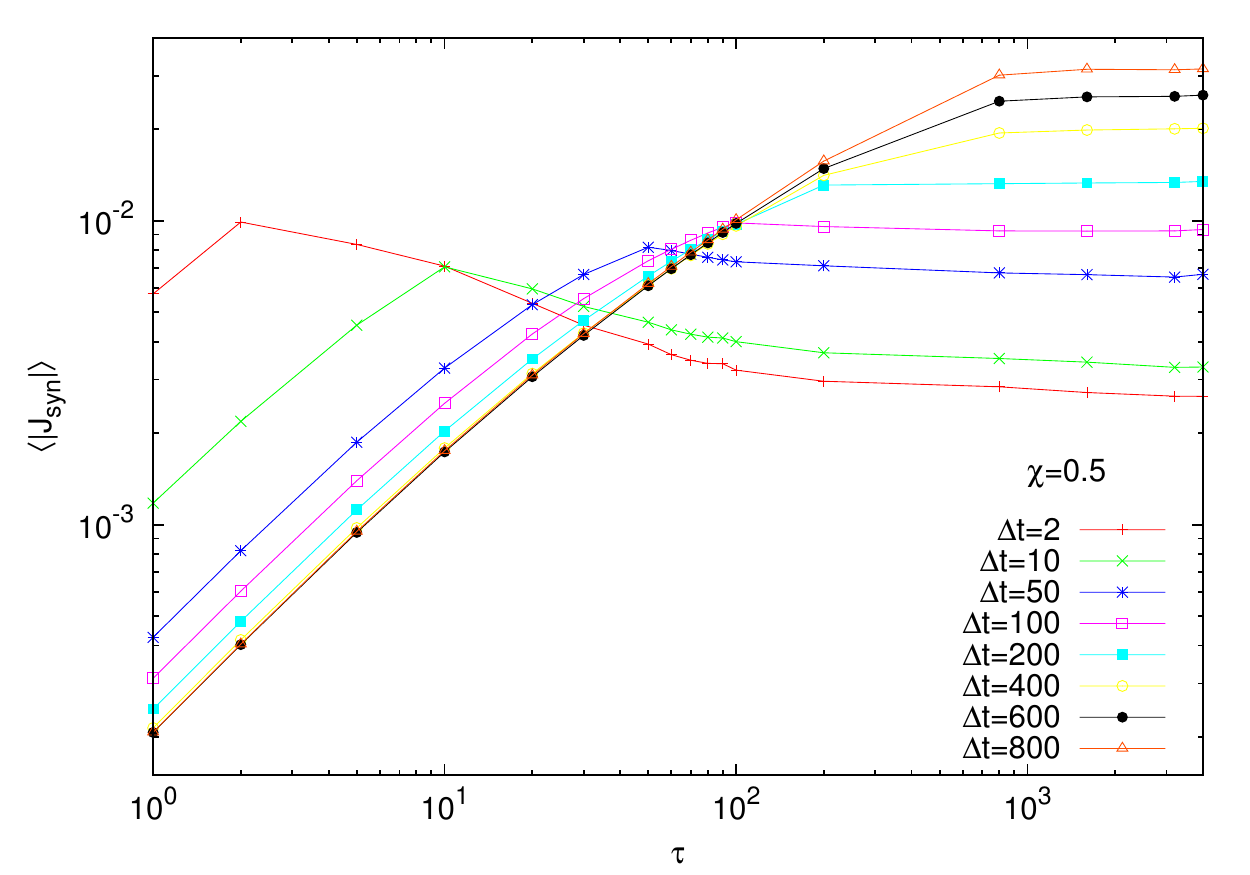}
        \caption{Upper panel: $\langle|J_{\text{eq}}|\rangle$ versus various time bin lengths $\Delta t$, for different values of $\chi$. Bottom panel: $\langle|J_{\text{sync}}|\rangle$ versus $\tau$ for various time bin lengths $\Delta t$, and $\chi=0.5$. With $\Delta t \leq 200$ seconds, $\langle|J_{\text{sync}}|\rangle$ has a maximum at $\Delta t = \tau$ , while for $\Delta t> 200$ seconds, $\langle|J_{\text{sync}}|\rangle$ tends to be a plateau.}
	\label{fig2}
      \end{figure}

Our mapping using volume information allows to obtain information on larger time scales, on which trade / no trade patterns are inefficient, because the window size is so large that always at least one trade happens during this time. Indeed, the local magnetizations do not necessarily go to one when the window size increases. As a consequence, the connected correlations do not tend to zero, and neither do the inferred couplings. This is illustrated by the top panel of figure \ref{fig2}, which shows that the average absolute value of the equilibrium couplings $\langle|J_{\text{eq}}|\rangle$, where $\langle.\rangle$ denotes the average over all pairs of stocks, increases with the window size, for different values of the volume threshold parameter $\chi$.  The evolution of the slope is however not monotonous with $\chi$. We study the absolute value of the couplings, because large negative values, which can counterbalance positive ones in the average without absolute value, might also give interesting information about the system.
Actually, for larger values of the threshold parameter, $\chi=2$ for instance, large negative values of couplings are an important contribution to the high value of $\langle|J_{\text{eq}}|\rangle$.
{\color{black}We can note also that the correlations observed are much bigger than the $N^{-1/2}=10^{-3}$ which would be observed for independent spins, with data spanning over $10^6$ seconds.}

The same analysis performed on the synchronous couplings on the bottom panel of figure \ref{fig2} shows that for small values of the window size $\Delta t$, the average absolute value of couplings has a maximum for a value of the time-lag of correlations $\tau=\Delta t$. This corresponds to the smallest time-lag which presents no redundancy between the two considered time windows. For higher values of the window size, big values of the time-lag give a maximal plateau.

\begin{figure}
  \centering
  \includegraphics[width=10cm]{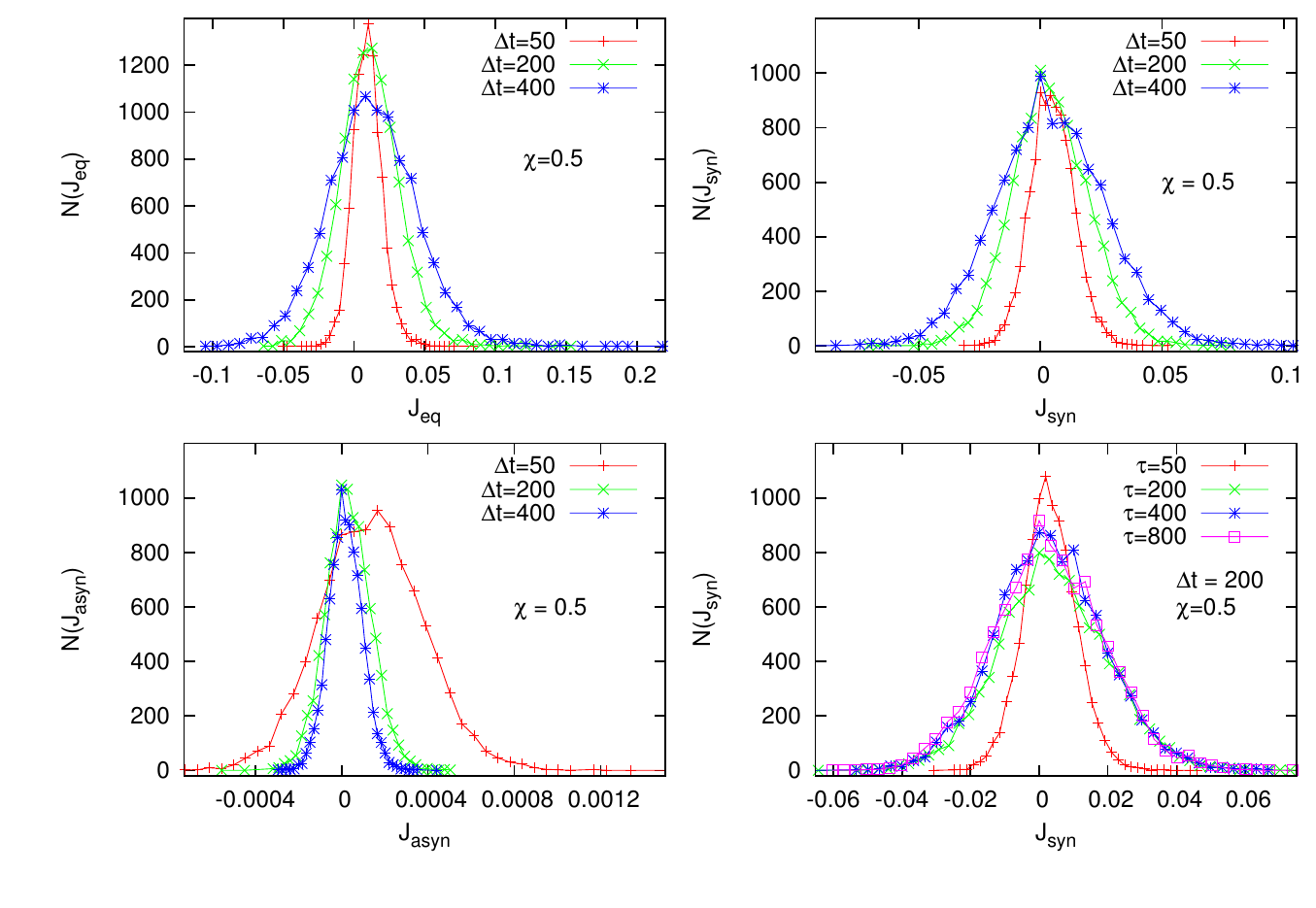}
  \includegraphics[width=10cm]{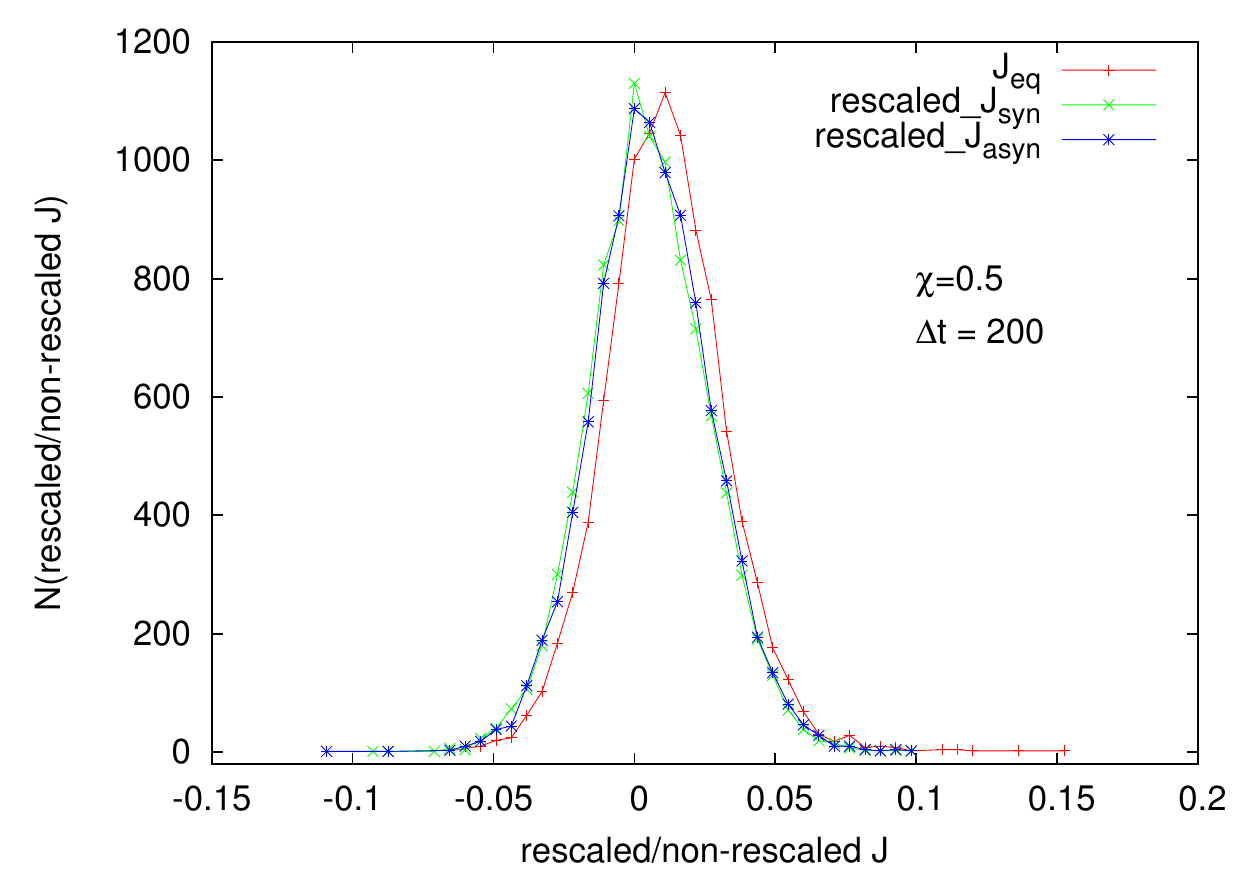}
  \caption{Histograms $N(J)$ of inferred couplings. Upper panel has four subplots: upper left one, histogram of $J_{eq}$ with different time bins, upper right $N(J_{syn})$, using $\tau = \Delta t$; bottom left $N(J_{asyn})$ and bottom right $N(J_{syn})$ with different values of $\tau$. Bottom panel: couplings obtained by the three inference methods. $J_{syn}$ and $J_{asyn}$ are rescaled to have the same standard deviation as $N(J_{eq})$.  For the three versions,  $\chi=0.5$ and $\Delta t= 200$ seconds, and for synchronous inference $\tau=\Delta t$. 
}\label{fig3}
\end{figure}

The goal of this study is to find values of the parameters introduced by the mapping and the inference, which yield inferred couplings containing interesting information. A first rough approach is to consider that couplings contain more information if they are bigger in absolute value. A more refined, but also laborious one, consists in looking at the distributions of couplings for different values of the parameters. This is done on figure \ref{fig3}. For asynchronous inference, the derivative of the time-lagged correlations $\dot C_{ij}(\tau)$ (see \ref{inference}) is computed through a linear fitting of this function $C_{ij}(\tau)$ using four points: $C(0)$, $C(\Delta t/5)$, $C(2\Delta t/5)$ and $C(3\Delta t/5)$. This explains why the histogram of $J_{asyn}$ becomes sharper when $\Delta t$ is increased on the upper panel of figure \ref{fig3}, as this parameter is then in the denominator of the derivative.

The bottom panel of figure \ref{fig3} shows that the three inference methods give close results when the distributions of couplings are considered.
For comparison, the distributions are rescaled on the bottom panel so as to have the same standard deviation. The upper panel shows how these distributions change with the parameters. It can be remarked that for small time scales, they have a strictly positive mean and a long positive tail. For higher time scales, the distributions are more centered around zero, but they keep an asymmetry and a longer positive tail than the negative one. This prevalence of positive couplings can intuitively be linked with the market mode phenomenon.

Indeed, a spectral analysis of the inferred coupling matrices (not presented here) yields the well-known fact (at least for correlations or couplings of stock prices \cite{bouchaud2003,mantegna1999,biely2008,bury2013}) that a large eigenvalue appears, corresponding to a collective activity of all stocks, usually named market mode. However, we did not manage to give an interpretation to the eigenvectors corresponding to the next eigenvalues in order of magnitude, as they do not seem to correspond to specific industrial or activity sectors for instance.

With increasing values of $\Delta t$, the histograms of $J_{eq}$ (and $J_{syn}$ with $\tau=\Delta t$) become broader, which is consistent with the increasing values of $\langle|J_{\text{sync}}|\rangle$ observed on the upper panel of figure \ref{fig2} (and with the lower panel of figure \ref{fig2}). The last figure of the upper panel of figure \ref{fig3} shows that the histogram of $J_{syn}$ does not change much with $\tau$ for high values of this parameter, which $\langle|J_{\text{sync}}|\rangle$ indicates also on the bottom panel of figure \ref{fig2}. 

\ref{similarity} describes a more precise measurement $Q_{JJ'}$ of the similarity between interactions matrices $J$ and $J'$ inferred by different methods.
\begin{figure}
  \centering
  \includegraphics[width=8cm]{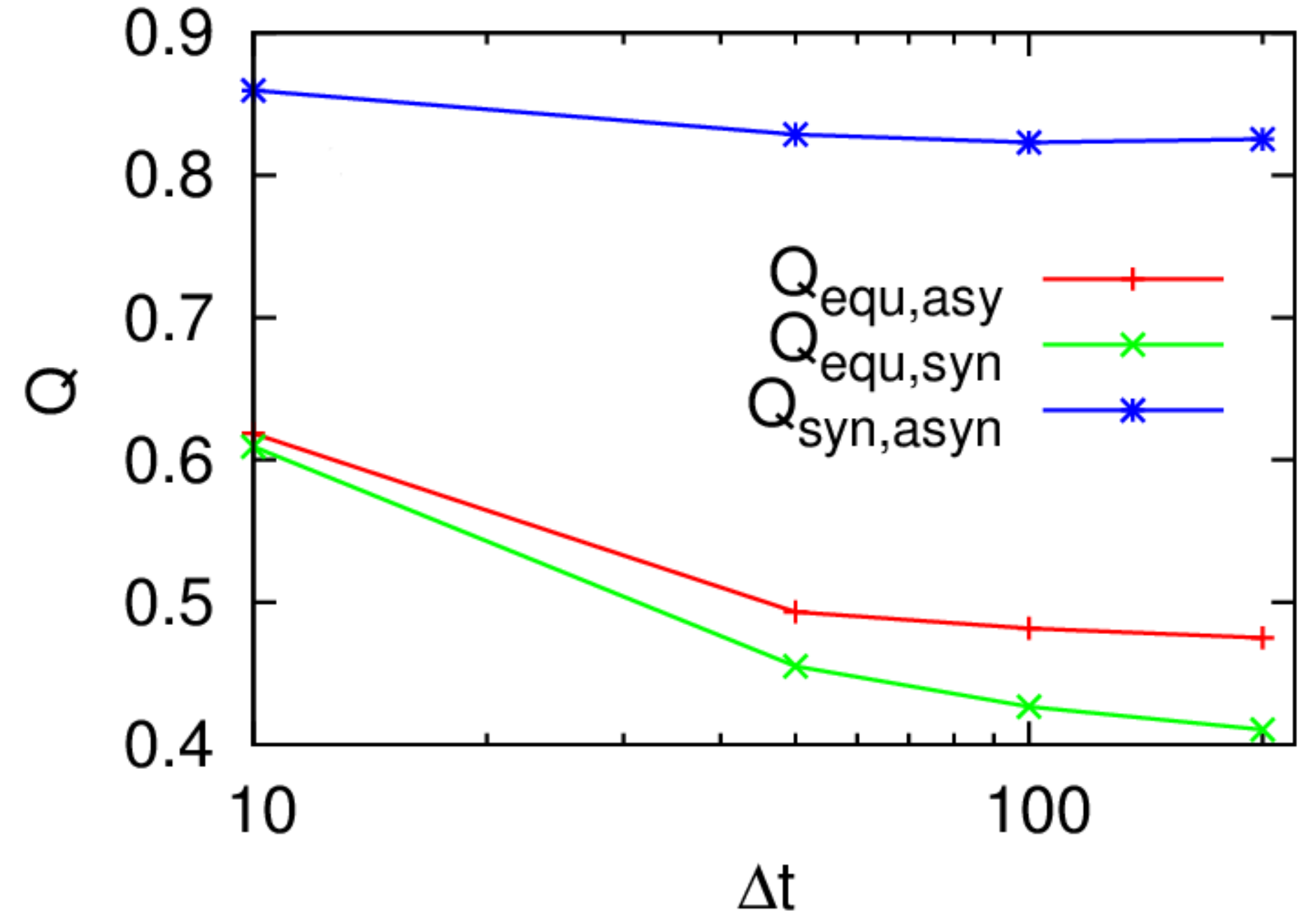}
  \caption{Similarity $Q_{JJ'}$ between interaction matrices obtained with different inference methods, versus window length $\Delta t$. The couplings are rescaled to have the same mean. $\chi=0.5$, $\tau=\Delta t$ for the synchronous inference, and the same fitting as for figure \ref{fig3} is used for the asynchronous inference. }\label{fig5b}
\end{figure}
Figures \ref{fig5b} (see also figure \ref{fig5a} in the Appendix) shows that the inferred coupling matrices with different methods are much more similar than matrices whose elements are drawn at random from the same distribution, which also indicates the coherence of the different inference methods. Indeed, figure \ref{fig5b} displays high similarities between couplings obtained from equilibrium, synchronous and asynchronous inference. Synchronous and asynchronous inference give especially close results, while equilibrium inference gives couplings which differ more from the other two methods. Let us note that synchronous and asynchronous methods infer directed networks, whereas equilibrium inference infers an undirected one. All similarities decrease when $\Delta t$ increases, which might indicate that the specificity of each method appears more clearly on bigger time scales.

\section{Financial networks}
\label{networks}
\begin{figure}[h!]
      \centering
\begin{tabular}{cc}
        \includegraphics[width=8.2cm]{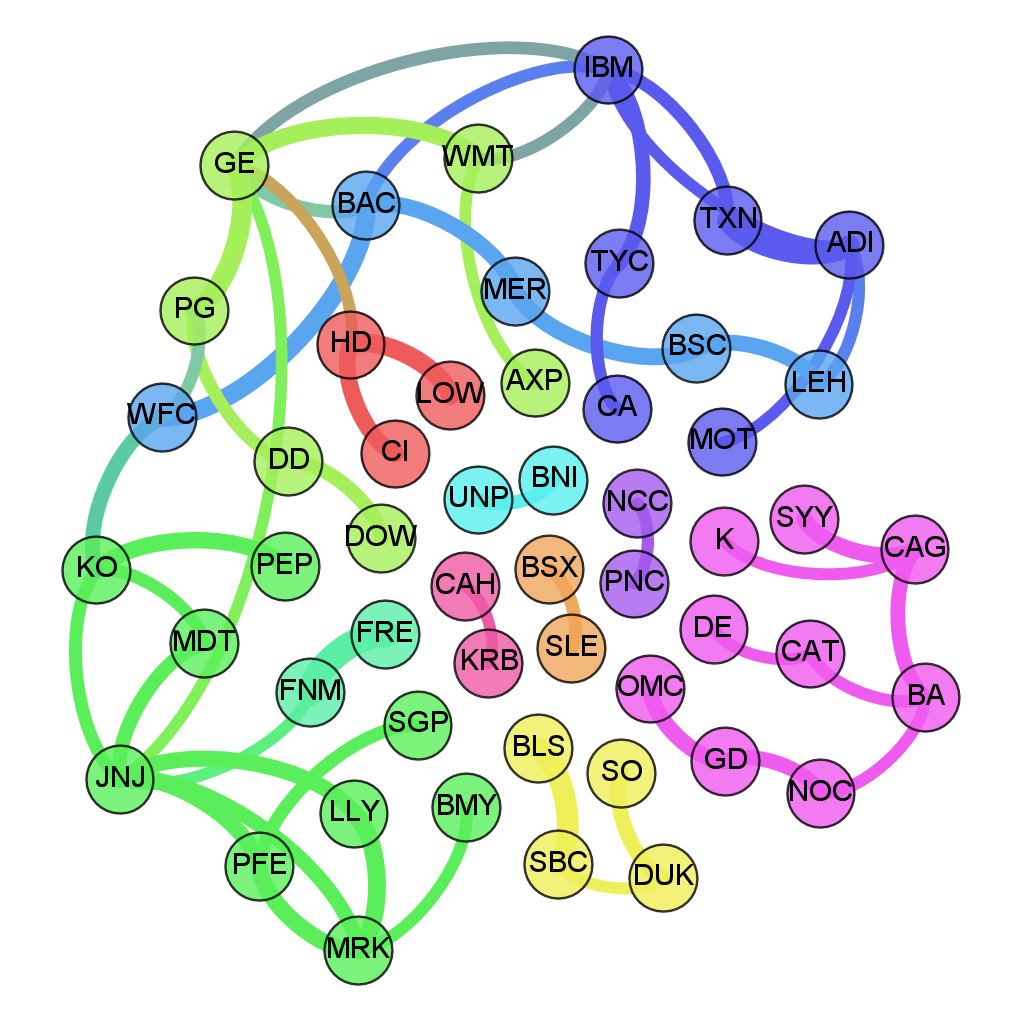}&
	\includegraphics[width=8.2cm]{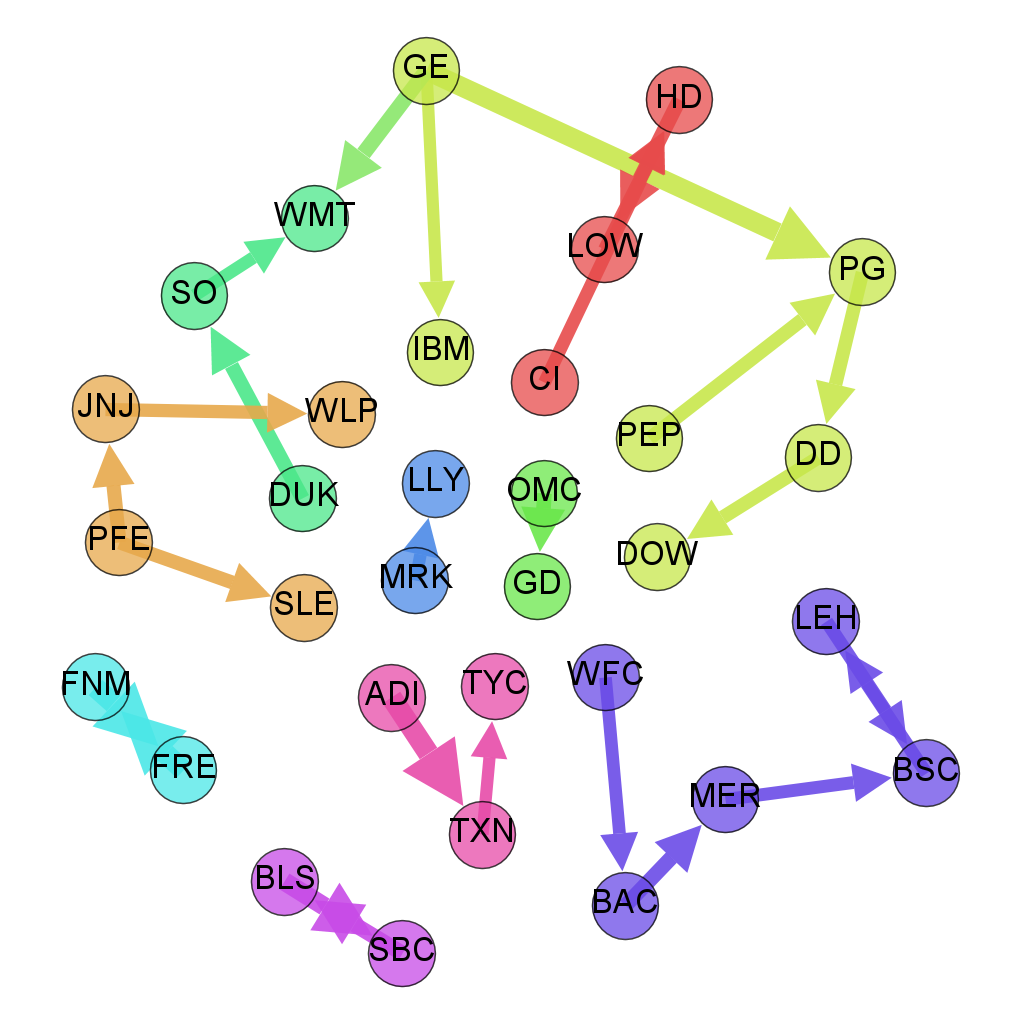}
\end{tabular}
        \caption{Inferred financial networks, showing only the largest interaction strengths (proportional to the width of links and arrows). Colors are indicative, and chosen by a modularity-based community detection algorithm \cite{blondel2008}. Parameters: $\chi=0.5$ and $\Delta t= 100$ seconds. Left panel: equilibrium inference. Right panel: synchronous inference, with $\tau=20$ seconds.}
	\label{fig5}
      \end{figure}

Although the spectral analysis does not distinguish industrial sectors, these can still be found in the inferred couplings, over a wide range of parameters, as figure \ref{fig5} shows. We focus only on the largest couplings, which can be easily explained by closely related activities of the considerd stocks. The left panel of figure \ref{fig5} shows that with equilibrium inference, more than half the stocks in the data can be displayed on a network where almost all links have simple explanations.

The following couplings in order of magnitude (not displayed on the figure) can also for some of them be given such simple explanations, but this is generally not true for small positive couplings. And negative ones do not actually give information which can be used easily, as they may simply indicate that the stocks are unrelated.

The network of the left panel of figure \ref{fig5} presents different communities, which are most of the time determined by a common industrial activity. Some of the links are very easy to explain with the proximity of activities (and often quite robust), for instance the pairs FNM - FRE (Fannie Mae - Freddie Mac, active in home loan and mortgage), UNP - BNI (Union Pacific Corporation - Burlington Northern Santa Fe Corporation, railroads), BLS - SBC (BellSouth - SBC Communications, two telecommunications companies now merged in AT\&T), NCC - PNC (National City Corp. - PNC Financial Services, now merged), HD - LOW (The Home Depot - Lowe's, both retailers of home improvement products), DOW - DD (Dow - DuPont, chemical companies),  MRK - PFE (Merck \& Co. - Pfizer, pharmaceutical companies), KO - PEP (The Coca-Cola Company - PepsiCo, beverages). 

These two last companies display strong links with the medical sector at different scales of volume and time, as KO here with MDT (Medtronic) and JNJ (Johnson \& Johnson). This medical sector is itself linked to the pharmaceutical sector with PFE, MRK, LLY (Lilly), BMY (Bristol-Myers Squibb) and SGP (Schering-Plough). Telecommunications (BLS, SBC) are linked to electric power with DUK (Duke Energy), itself linked to electric utilities with SO (Southern Company).

GE (General Electrics) is for a large range of parameters a very central node, which is consistent with its very diversified activities. On the left panel of figure \ref{fig5}, the relation between PG (Procter \& Gamble) and WMT (Walmart), both retailers of consumer goods, comes at this level of interaction strength through GE.

The banking sector, with WFC (Wells Fargo), BAC (Bank of America), MER (Merrill Lynch), BSC (Bear Stearns), LEH (Lehman Brothers) has here the structure of a chain. It is linked to the sector of electronic technology: IBM, TYC (Tyco International), CA (CA Technologies, software company), TXN (Texas Instruments), ADI (Analog Devices), MOT (Motorola, telecommunications).

The defense and aerospace sector, with GD (General Dynamics), NOC (Northrop Grumman) and BA (Boeing), is linked to engines and machinery with CAT (Caterpillar Inc.) and DE (John Deere), and more strangely, to packaged food with CAG (ConAgra Foods), SYY (Sysco) and K (Kellogg Company).

Some non-intuitive links appear on different time and volume scales and with different inference methods: the links of KO with the medical sector have been mentioned, DD is linked to PG, and GD is linked here with OMC (Omnicom Group, advertising and marketing), and sometimes with UNP or HDI (Harley-Davidson, motorcycles manufacturer).

The right panel of figure \ref{fig5} presents the results of synchronous inference in the same conditions. It shows that the results of equilibrium and synchronous inference are consistent, and that synchronous inference provides additional information, as it infers an undirected network (all this is also true for asynchronous inference). For instance, GE is clearly a node which influences others and is not strongly influenced itself at this level of interaction, and the financial sector is a directed chain. Stocks whose tickers are not indicated in the text are described in \ref{stocks}.

To sum it up, what do these results tell about financial markets? First, they go in the same direction as the results about market mode \cite{bouchaud2003,mantegna1999}: most of the interaction strengths found are usually positive, which indicates that the financial market has a clear collective behaviour, even when only trade and volume information is considered. Stocks tend to be traded or not traded at the same time, which can be partly explained by the fact evoked previously that the frequency spectrum of the data has a strong component corresponding to a collective motion with a frequency of one day. 

However, this does not explain the fact that the strongest inferred interactions can be easily understood by similarities in the industrial activities of the considered stocks. This means that financial activity -- a priori irrespectively of bullish or bearish market, as no price information is considered -- tends to concentrate on a certain activity sector at a certain time. For price dynamics this phenomenon is well-known \cite{biely2008,bury2013,kullmann2002}, but it is more surprising that it appears also on traded volumes.

Actually, these two phenomena (collective behaviour, and strong interactions within industrial sectors), and models such as \cite{cont2000} also support the idea that financial markets relie a lot on imitation, a fundamental social behaviour which is still very much at work in the 21st century society, as illustrated for instance by fashion trends, important news subjects, or even popular research themes. As a consequence, assessing the predictive power of such inferred models, at least on short time scales, is an interesting perspective of work.

\section*{Conclusion and perspectives}
Using a mapping of complex data to binary strings, we infer Ising models on financial data, based only on traded volume information. We recover standard results of the literature (obtained using price information), and the inferred networks give a consistent picture of the financial markets. Synchronous and asynchronous inference give a directed network in agreement with the undirected equilibrium one. An obvious first perspective is then to combine this volume information to price information in a more involved mapping (which could be inspired from the relationship between price and volume \cite{karpoff1987}), to obtain new insights about the interactions on the stock market. Indeed, price is usually more studied, because it corresponds essentially to the economic value of the considered stock on the market.

More immediate perspectives would be to study the influence of the time scale parameter $\tau$ of synchronous and asynchronous inference, and also to study close mappings using only volume information. Indeed, instead of using the average volume traded per unit of time as a reference, other variables characterizing the stock could be used, in order to obtain different information, for instance an indicator of the economic size of the considered company, such as revenue, operating income or total assets.

Other immediate possibilities would be to introduce a regularization ($L^1$ for instance), which would be a clearer way than thresholding to determine which interactions are meaningful. And also, to overcome a limit of our method, which is that for big scales of both volume and time, some stocks may have a -1 average magnetization (meaning that no traded volume exceeded the threshold on the whole period), which prevents us from inverting the correlation matrix and doing the inference. A solution to explore as a perspective of this work would be to take these stocks off the data, because they are inactive on these time and volume scales. It would even make sense to also ignore all stocks which do not have significant activity on these scales.

Finally, to test the results of different inference methods, the predictive power of such models (assessed for instance by measuring the distance between data generated by the model and real data following the period on which inference was done) on different time and volume scales could be a valuable indicator of their worth. With additional price information, this predictive power can be expected to be greater. It is also an interesting perspective from a financial engineering point of view, with different goals.

\subsection*{Acknowledgements}
The authors thank Matteo Marsili for providing the data, and acknowledge interesting discussions with Erik Aurell, Matteo Marsili and Iacopo Mastromatteo. This work was supported by funding from the Center of Excellence program of the Academy of Finland, for the COMP Center.

\section*{References}
\bibliographystyle{unsrt}
\bibliography{NYSE}

\appendix
\section{Inference methods}
\label{inference}
The mapping described in section \ref{data} provides us with time series $s_i(t)$ for each stock. With this, it is natural to define local magnetizations $m_i$ and correlations $C_{ij}(\tau)$ as follows:
\[
m_i=\langle s_i(t)\rangle_t
\text{ and }
C_{ij}(\tau)=\langle s_i(t+\tau)s_j(t)\rangle_t-m_im_j
\]
where $\langle .\rangle_t$ denotes time averaging over the whole data length.

Several inference methods have been developed recently. We focus here on mean-field approximation, using three different versions, which can be described as follows. 
Let us first present the inference formula for coupling strengths, which differs for each method:
\begin{itemize}
  \item Equilibrium inference, which focuses on equal time correlations \cite{Kappen1998}
\[
J_{ij} = \frac{\delta_{ij}}{1-m_i^2} - C(0)_{ij}^{-1}
\]

  \item Synchronous inference is suitable for non-equilibrium inference, and considers also time-lagged correlations with a time lag $\tau$ in addition to equal time correlations \cite{roudi2011}:
    \[\label{Syn}
    J_{ij} = \frac{1}{1-m_i^2}(C(\tau)C(0)^{-1})_{ij}
    \]

  \item Asynchronous inference, also modeling non-equilibrium processes, uses the derivative of the time-lagged correlations $\dot C_{ij}(\tau)$ \cite{zeng2013}:
      \[\label{Asyn}
        J_{ij} = \frac{1}{1-m_i^2}\Big(\frac{dC(\tau)}{d\tau}|_{\tau=0}.C(0)^{-1}\Big)_{ij}
      \]
\end{itemize}
And the inference formula for fields is each time the same:
\[
h_i = \texttt{atanh}~m_i - \sum_{j\neq i} J_{ij}m_j.
\]
The main difference between the two last methods is that synchronous inference assumes that all spins are proposed to update in parallel at discrete time instances, while asynchronous inference does not have a such assumption: update times themselves are stochastic variables. The asynchronous method is supposedly more powerful, as it monitors the decay in time of all pair correlations, and thus the time derivative $\dot C_{ij}(\tau)$ appears in the formula.

It can be remarked that, in addition to the mapping parameters $\Delta T$ and $\chi$, the synchronous and asynchronous inference methods have a parameter $\tau$, which is in these respective cases the considered time-lag of correlations, and the time scale on which the derivative of correlations $dC(\tau)/d\tau|_{\tau=0}$ is computed. For the asynchronous case, this time scale does not appear explicitly in the formula, but arises when the derivative is computed from the data.

\section{Similarities}
\label{similarity}
We want to study more precisely the closeness between couplings obtained with different methods. The distributions presented on figure \ref{fig3} give a first idea, but we want to assess how similar individual couplings $J_{ij}$ and $J'_{ij}$ are. To this end, we introduce for two coupling matrices $J_{ij}$ and $J'_{ij}$ a similarity measure $Q_{JJ'}$ given by
\[
Q_{JJ'}=\frac{\sum_{i,j}J_{ij}J'_{ij}}{\sum_{i,j}\max (J_{ij},J'_{ij})^2}
\]
This measure considers actually these coupling matrices as the vectors containing all their elements, compares them one by one and gives a global similarity measure. It takes real values between 1 (when $J_{ij}=J'_{ij}$ for all $i$ and $j$) and -1 (when $J_{ij}=-J'_{ij}$ for all $i$ and $j$), and values close to zero indicate uncorrelated couplings.
\begin{figure}
  \centering
  \includegraphics[width=8cm]{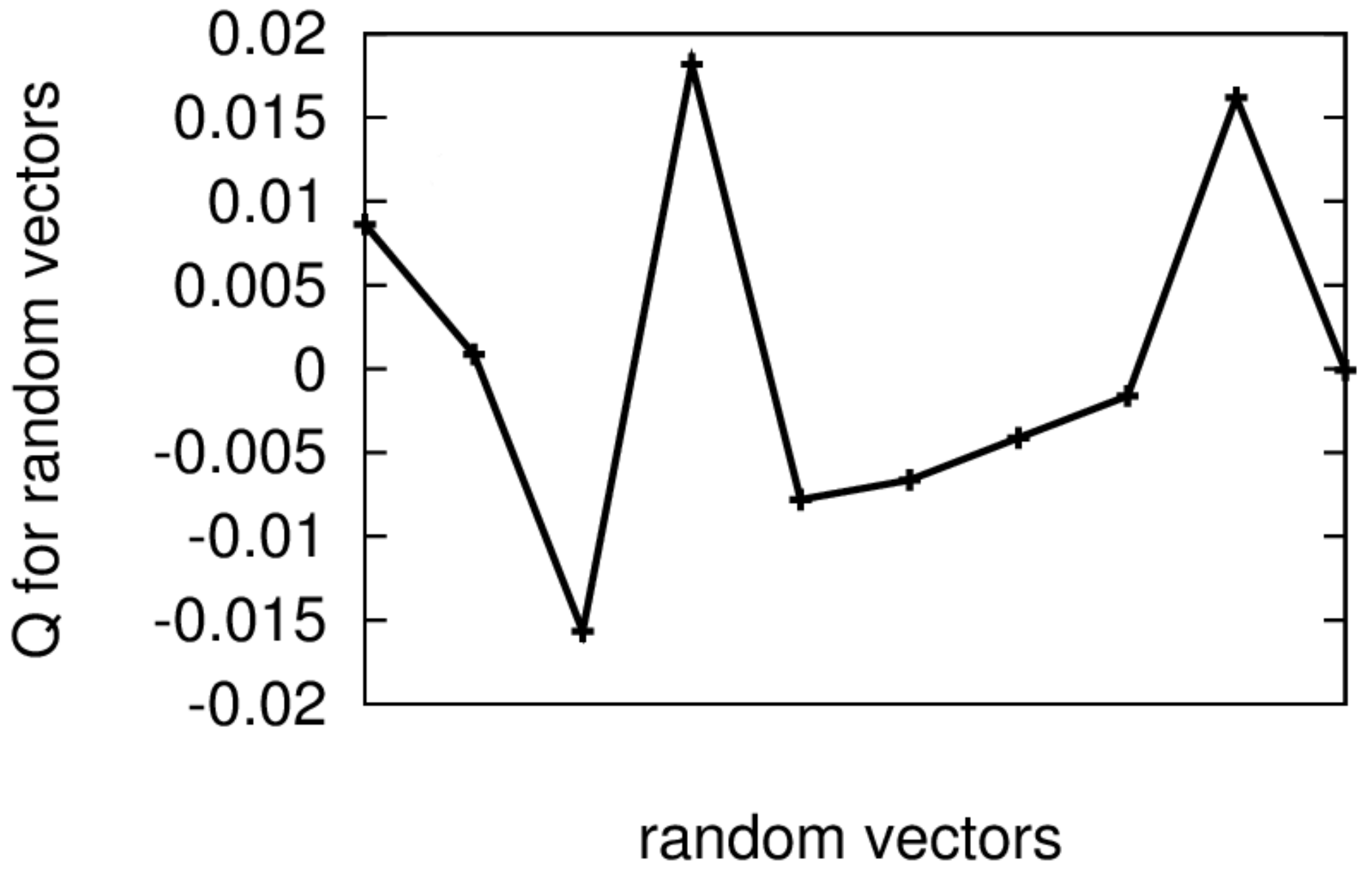}
  \caption{Similarity $Q_{JJ'}$ between random matrices whose elements are drawn independently from the same normal distribution. This distribution has mean $0$ in each case, and standard deviation $1,2,...,10$ for successive points from left to right.}\label{fig5a}
\end{figure}
For comparison with the similarity measures obtained between two different methods on figure \ref{fig5b}, we show on figure \ref{fig5a} that this similarity measure takes small values (smaller than $0.02$ in absolute value) when all elements of the vectors $J_{ij}$ and $J'_{ij}$ are drawn independently at random from the same Gaussian distribution, of mean 0, and for different values of the standard deviation of this distribution. Negative values are even obtained sometimes, as can be expected from the formula defining $Q_{JJ'}$. However, figure \ref{fig5b} displays much bigger similarities between the different inference methods.

\section{Description of stocks}
\label{stocks}
\begin{table}[h!]
\centering
\begin{tabular}{|l|l|l|}
\hline
Ticker & Name & Activity\\
\hline
AXP & American Express & Financial services\\
BSX & Boston Scientific & Medical devices\\
CAH & Cardinal Health & Pharmaceutical and medical products\\
CI & Cigna & Health care management\\
KRB & MBNA & Banking\\
SLE & Sara Lee Corporation & Consumer-goods\\
WLP & WellPoint & Health care management\\
\hline
\end{tabular}
\caption{\label{stocks1} {Description of the stocks whose tickers are not mentioned in the text.} }
\end{table}

\end{document}